\documentclass[prl,twocolumn]{revtex4}
\usepackage{bm}
\usepackage{graphicx}
\usepackage{amssymb}
\usepackage{amsmath}
\usepackage{eufrak}

\begin{document}

\title{Magneto-Photoluminescence of InAs/InGaAs/InAlAs quantum well structures}

\author{Ya.\,V.~Terent'ev$^{1,2}$, S.\,N.~Danilov$^1$, J.~Loher$^1$,
D.~Schuh$^1$, D.~Bougeard$^1$, D.~Weiss$^1$, M.\,V.~Durnev$^2$,
S.\,A.~Tarasenko$^2$, M.\,S.~Mukhin$^2$, S.\,V.~Ivanov$^2$, and
S.\,D.~Ganichev$^1$}
\affiliation{$^1$ Physics Department, University of Regensburg,
93040 Regensburg, Germany}
\affiliation{$^2$Ioffe Physical-Technical Institute,
194021 St.~Petersburg, Russia}
\begin{abstract}
Photoluminescence (PL) and highly circularly-polarized magneto-PL (up to 50\% at 6 T) from
two-step bandgap InAs/InGaAs/InAlAs  quantum wells (QWs) are
studied. Bright PL is observed up to room temperature, indicating
a high quantum efficiency of the radiative recombination in these
QW. The sign of the circular polarization indicates that it  stems
from the spin polarization of heavy holes caused by  the Zeeman
effect. Although in magnetic field the PL line are strongly
circularly polarized, no energy shift between the counter-polarized PL lines was observed.
The results suggest that the electron and the hole
$g$-factor to be of the same sign and close magnitudes.
\end{abstract}

\maketitle{}

Narrow bandgap InAs-based heterostructures, being characterized by
a high carrier mobility and a strong spin-orbit interaction, are
generally addressed as promising systems for high-frequency
electronics, optoelectronics and spintronics. In particular, QWs
based on a InAs/InGaAs/InAlAs two-step bandgap engineering
offer optimized confinement properties. Such QWs have been used to
demonstrate a broad tunability of the emission energy in the
mid-infrared range~\cite{Tournie1992},
high quality two-dimensional  carrier
systems~\cite{Richter2000,Heyn2003,Moller2003,Hirmer2011},
pronounced spin phenomena~\cite{Hu2003,Wurstbauer2009,Wurstbauer2010,Olbrich2012},
and an electrical tunability of the electron
$g$-factor~\cite{Nitta2}. A precise knowledge of the carrier $g$-factors will be a key to evaluate the application potential of
such systems in spintronics. Currently, reported values of
electron $g$-factors determined in InAs/InGaAs/InAlAs QW by
magneto-transport and terahertz experiments vary in a broad range
from -3.1 to -9 depending on
the In-content of the QW~\cite{Moller2003,Hu2003,Nitta2},
while $g$-factors deduced from
magneto-optical spectroscopy are found to be much
smaller in magnitude~\cite{Tsumura2006}.
In this letter, we address the discrepancy in observed $g$-factors
by reporting on polarization-resolved magneto-PL of samples with a high In content,
for which high electron $g$-factors have been reported~\cite{Hu2003}.
We indeed observe a strong PL polarization in a magnetic field, indicating large carrier
$g$-factors and leading to a degree of circular PL
polarization of up to 50\% at 6 T. We demonstrate that the spin polarization
of the heavy holes determines the observed PL polarization.
Interestingly, although the carrier $g$-factors
are found to be high, we do not observe any energy shift between the counter-polarized PL lines, resulting in a vanishing effective electron-hole $g$-factor.
From this observation, we conclude the electron and the hole $g$-factor to be
of the same sign and magnitude. These findings are conserved when varying
the InAs QW width.

\begin{figure}[b]
\includegraphics[width=0.6\linewidth]{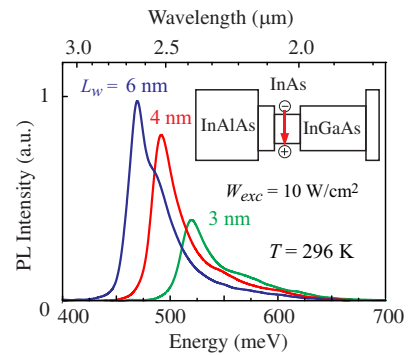}
\caption{PL spectra of undoped InAs QW samples.
The inset shows the band diagram of the active region.}
\label{fig01}
\end{figure}

The active region was grown by molecular beam epitaxy onto a fully
relaxed In$_x$Al$_{1-x}$As/(001)GaAs graded
buffer~\cite{Capotondi2005}
with a stepwise increase of the In content ($x$ = 0.05 to $x$ =
0.75) over 1~$\mu$m  and consisted of a single QW as sketched in
the inset of Fig.~\ref{fig01}. The QW potential barriers are built
from In$_{0.75}$Al$_{0.25}$As. To optimize the carrier confinement
within the QW, the core of the QW, a pure InAs layer is
asymmetrically embedded into a 16~nm thick
In$_{0.75}$Ga$_{0.25}$As. To further tailor the bandstructure and
hence the $g$-factor, different widths of the InAs layer of $L_w=$
3, 4 and 6~nm were used. An additional structure was modulation
doped with Si below the QW at a distance of 7.5~nm from the border
of the QW. The doping
induces a two-dimensional electron gas (2DEG)  in the QW region.
An electron density of $1 \cdot 10^{12}$~cm$^{-2}$ and a mobility of
$7\cdot 10^4$~cm$^2$/(V$\cdot$s) at $T=14$~K was determined in
magneto-transport experiments.  To demonstrate the resolution of
our magneto PL  set-up an additional In$_{x}$Al$_{1-x}$As QW
structure with Mn delta-layer in the barrier has been
prepared~\cite{Wurstbauer2009,Olbrich2012}.

PL detected with a Fourier Transform Infrared
(FTIR) spectrometer was excited by a laser diode operating in the
\textit{cw} mode at wavelength $\lambda = 809$~nm. The laser beam
was focused to a 1-mm diameter spot on the sample. The
excitation density $W_{exc}$ was varied from 0.5 to 20~W/cm$^2$.
An external magnetic field up to 6~T was applied perpendicularly
to the wafer along the detected emission (Faraday geometry).
The sample temperature was varied from 2 to
300~K. Right- and left-handed circular polarized emission spectra
were recorded applying a quarter wave ZnSe Fresnel
rhomb~\cite{GanichevPrettl}.

Bright room-temperature PL is observed from all the undoped
samples, see Fig.~\ref{fig01}. Depending on the QW width, the PL
spectral position varies from $\lambda = $ 2.4 to 2.65~$\mu$m.
According to calculations, this corresponds to direct optical
transitions between the ground electron  $e1$ and the heavy
hole $hh1$ subbands. The PL spectral width is about 30~meV, being
close to the thermal carrier energy at room temperature. By
cooling the samples down to 2~K, the PL intensity increases by a
factor of 20 and the linewidth decreases to 20~meV. In the whole
range of excitation densities used, the PL intensity linearly
depends on the excitation power, indicating a high quantum
efficiency of radiative recombination in the QW structures.

\begin{figure}[t]
\includegraphics[width=0.6\linewidth]{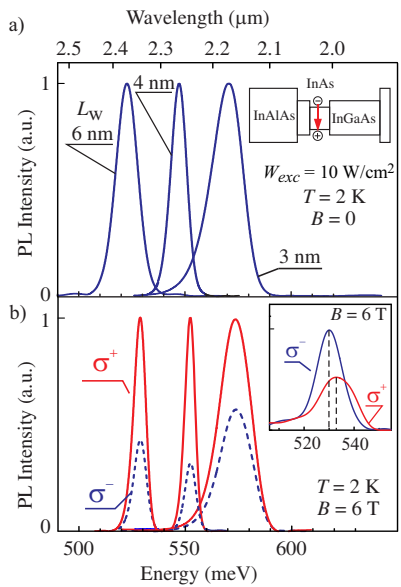}
\caption{
(a) PL spectra of undoped InAs QWs.
(b) Circularly polarized magneto-PL spectra of the same QWs.
Inset shows the PL spectra from a reference structure doped with Mn.
}
\label{fig02}
\end{figure}

The application of a magnetic field $B$
leads to a circular polarization of the PL, as shown in
Fig~\ref{fig02}b for undoped samples, with a predominant emission
of $\sigma^+$ photons along the field direction. The degree of
circular polarization $P_{circ} = (I_+ - I_-)/(I_+ + I_-)$, where
$I_{+/-}$ is the intensity of the $\sigma^{+/-}$ circularly
polarized emission, is constant within the emission spectrum and
exceeds 50\% at $T = 2$~K and $B=$~$6$~T. Surprisingly, despite
the strong circular polarization and the narrow emission lines, no
energy splitting of the circularly polarized components is
observed for all samples studied. Note, that a splitting down to
$0.3$~meV could be safely detected in our setup as it was 
observed, e.g., in a weakly Mn-doped
reference samples of similar design,
see inset in Fig.~\ref{fig02}b.

\begin{figure}[t]
\includegraphics[width=0.6\linewidth]{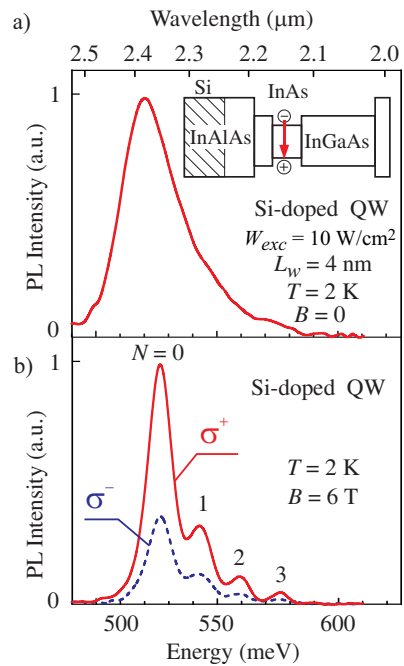}
\caption{
(a) PL spectra of Si-doped InAs QW.
(b) Circularly polarized magneto-PL spectra.
The numbers $N$ denote the transitions with the corresponding Landau level numbers.
}
\label{fig03}
\end{figure}

PL is also detected from the heavily $n$-doped QW structure
containing a 2DEG, Fig.~\ref{fig03}a. Its intensity is about one order of magnitude
weaker than the one observed from the undoped samples.
Furthermore, at zero magnetic field, the PL spectrum is substantially wider and
has an asymmetric shape.
The linewidth corresponds approximately to the electron Fermi energy $E_F = 60$ meV. The
latter value is obtained from the measured  2DEG density, $1
\times 10^{12}$~cm$^{-2}$, and the effective electron mass $m_e^*
=0.038 m_0$, determined by magnetotransport measurements in
similar structures~[\onlinecite{Richter2000,Moller2003,Hu2003}].
The application of an external magnetic field leads to a
multi-peak structure of the emission spectrum, which is caused by
the formation of Landau levels, see Fig.~\ref{fig03}b.
The data reveal that the degree of PL circular polarization is almost the same
for optical transitions corresponding to the
Landau level number $N=0,1,2$ and no energy shifts between the 
counter-polarized PL lines are detected.
The PL polarization degree reaches 50~\% at low temperatures and
$B=6$~T and is completely determined by the hole spin polarization
since the electron spin levels below the Fermi energy are equally
occupied. The energy distance between adjacent peaks is given by
$e \hbar  B /\mu$, where $\mu
= m_e^* m_{hh}^* / (m_e^* + m_{hh}^*)$ is the reduced mass, and
$m_{hh}^*$ is the in-plane mass in the heavy-hole subband. From
the data of Fig.~\ref{fig03} we find $\mu \approx 0.039 m_0$ which
is close to the effective electron mass
\cite{Richter2000,Moller2003,Hu2003}.

\begin{figure}[t]
\includegraphics[width=0.5\linewidth]{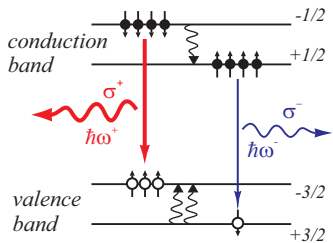}
\caption{Energy bands in magnetic field and allowed optical
transitions. The vertical arrows indicate optical transitions with
the emission of $\sigma^+$-
and $\sigma^-$-polarized
photons. Vertical waved arrows sketch the spin relaxation process.
Electron and hole population are shown schematically by circles.
}
\label{fig04}
\end{figure}

The observation that $\sigma^+$-polarized emission dominates in both
undoped and heavily $n$-doped samples indicates that the PL polarization 
results from the spin polarization of holes in the magnetic field.
This is particularly clear for the doped sample where both electron spin
levels below the Fermi energy are equally occupied. It is also true
for the undoped samples because the electron $g$-factor is known to be negative in InAs layers
($g_e \approx -15$ in bulk InAs~\cite{InAsgfactor} and $g_e \approx -7$ as we estimate for
In$_{0.75}$Ga$_{0.25}$As). Indeed, in the case of PL polarization dominated by the electron spin polarization, 
the PL helicity would have to be opposite to the observed one.
The dominating $\sigma^+$ circular polarization of the PL also
indicates that the heavy hole $g$-factor is negative and that $|hh1,-3/2 \rangle$
is the ground spin subband, as shown in Fig.~\ref{fig04} and expected from theory~\cite{Winkler,Durnev2014}.
The preferential population of the ground spin subband sketched in Fig.~\ref{fig04} is due to
the fast spin relaxation of photoexcited holes.
To explain the absence of a shift between the counter-polarized PL lines, we suggest that the Zeeman
splittings of the electron and hole states are similar.
In this case, the energies of the allowed
optical transitions~\cite{OO_book} $|e1,+1/2 \rangle \rightarrow |hh1,+3/2
\rangle$ and $|e1,-1/2 \rangle \rightarrow |hh1,-3/2 \rangle$
corresponding to $\sigma^-$- and $\sigma^+$-polarized photons,
respectively, are approximately the same.
To explain the absence of a shift between the counter-polarized PL lines, we suggest that the Zeeman
splittings of the electron and hole states are similar.
In this case, the energies of the allowed
optical transitions $|e1,+1/2 \rangle \rightarrow |hh1,+3/2
\rangle$ and $|e1,-1/2 \rangle \rightarrow |hh1,-3/2 \rangle$
corresponding to $\sigma^-$- and $\sigma^+$-polarized photons,
respectively, are approximately the same.

\begin{figure}[t]
\includegraphics[width=0.9\linewidth]{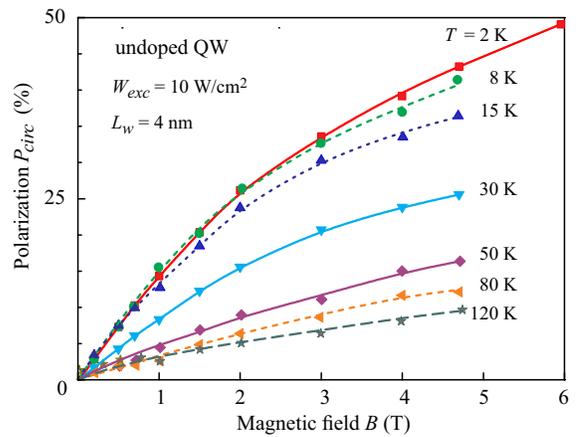}
\caption{ Magnetic field dependencies of PL polarization
measured in undoped QW.
Lines are the guide for eyes. }
\label{fig05}
\end{figure}

Figure~\ref{fig05} shows the magnetic field dependence of
$P_{circ}$ obtained
for one of the undoped samples. The polarization
increases linearly at low fields, $B \leq 2$~T and tends to
saturate at higher magnetic fields. With decreasing temperature,
the polarization considerably grows, reaching
50\% at $B = 6$~T  at $T = 2$~K.
This behavior is in line with the
microscopic model proposed above: $P_{circ}$ is proportional to
the spin polarization of holes, which is determined by $g_{hh}
\mu_B B/ (k_B T)$ at low magnetic fields. $g_{hh}$ is the hole
$g$-factor and $\mu_B$ the Bohr magneton. At higher fields, the polarization tends to saturate.
The degree of the electron and hole spin-polarization depends on the
respective spin relaxation, which leads to different population
of the spin split subbands, and recombination processes. We note
that at low temperatures, $T < 15$~K, the PL spectra recorded for
both circular polarizations are weakly sensitive to the
temperature, especially at low magnetic fields, see
Fig.~\ref{fig05}. This indicates that the carriers are efficiently
heated by radiation.

To summarize, the observed bright PL in 2.5~$\mu m$
spectral range from InAs QW structures shows that these
structures are of particular importance for optoelectronics application in
the mid-infrared range.
Strong circular polarization of the magneto-PL together with the absence of
the  line spectal splitting
reveal that the $g$-factors of electrons and heavy holes while
being large have the same sign and are close to each other in magnitude.

We acknowledge financial support from the DFG (SFB~689),
BMBF,
RFBR,
RF President Grant MD-3098.2014.2,
and ``Dynasty'' Foundation.


\begin{thebibliography}{99}

\bibitem{Tournie1992} E.~Tourni\'{e}, O.~Brandt, and K.~Ploog, Appl. Phys. Lett. \textbf{60}, 2877 (1992).

\bibitem{Richter2000} A. Richter, M. Koch, T. Matsuyama, Ch. Heyn, and U. Merkt,
Appl. Phys. Lett. \textbf{77},  3227
(2000).

\bibitem{Heyn2003} Ch. Heyn, S. Mendach, S. L\"ohr, S. Beyer, S. Schn\"{u}ll, and W. Hansen,
J. Crist. Growth \textbf{251},  832
(2003).

\bibitem{Moller2003} C. H.~M\"oller, Ch. Heyn, and D. Grundler, Appl. Phys. Lett. \textbf{83},  2181 (2003).

\bibitem{Hirmer2011} M. Hirmer, D. Schuh, and W. Wegscheider, Appl. Phys. Lett. \textbf{98}, 082103 (2011).

\bibitem{Hu2003} C.-M. Hu, C. Zehnder, Ch. Heyn, and D. Heitmann,
Phys. Rev. B \textbf{67}, 201302(R) (2003).

\bibitem{Wurstbauer2009}  U.~Wurstbauer, M.~Soda, R.~Jakiela, D.~Schuh, D.~Weiss, J.~Zweck, W.~Wegscheider,
J. of Cryst. Growth 311, 2160
(2009).

\bibitem{Wurstbauer2010} U.~Wurstbauer, S.~Sliwa, D.~Weiss, T.~Dietle, and W.~Wegscheider,
Nature Physics \textbf{6}, 955
(2010).

\bibitem{Olbrich2012} 
P. Olbrich, C. Zoth, P. Lutz, C. Drexler, V. V. Bel'kov,
Ya.~V.~Terent'ev, S. A. Tarasenko, A. N. Semenov, S. V. Ivanov, D.
R. Yakovlev, T. Wojtowicz, U. Wurstbauer, D. Schuh, and
S.~D.~Ganichev,
Phys. Rev. B \textbf{85}, 085310 (2012).


\bibitem{Nitta2}  
J. Nitta, Y. Lin, T. Akazaki, and T. Koga,
Appl. Phys. Lett. \textbf{83}, 4565 (2003).

\bibitem{Tsumura2006} 
K. Tsumura, S. Nomura, T. Akazaki, and J. Nitta, Phys. E
\textbf{34}, 315 (2006).

\bibitem{Capotondi2005} 
F. Capotondi, G. Biasiol, D. Ercolani, V. Grillo, E. Carlino, F.
Romanato, and L. Sorba, Thin Solid Films \textbf{484}, 400 (2005).

\bibitem{GanichevPrettl} S.D.~Ganichev and W.~Prettl,
\textit{Intense Terahertz Excitation of Semiconductors} (Oxford
Univ. Press, 2006)


\bibitem{InAsgfactor}
J. Konopka, Phys. Lett. A \textbf{26}, 21 (1967).

\bibitem{Winkler}
%
R.~Winkler,  \textit{Spin-Orbit Coupling Effects in Two-Dimensional Electron and Hole Systems},
(Springer
2003).

\bibitem{Durnev2014} M. V. Durnev, Phys. Solid State {\bf 56},  (2014).

\bibitem{OO_book} {\it Optical Orientation}, edited by F. Meier and B. P. Zakharchenya (Esevier Science, Amsterdam, 1984).



\end{thebibliography}
\end{document}